\documentclass[12pt]{iopart}

\usepackage{graphicx}
\usepackage{cite}
\begin{document}

\title{Multi loop soliton solutions and their interactions in the Degasperis-Procesi equation}

\author{S Stalin$^1$ and M Senthilvelan$^1$}

\address{$^1$ Centre for Nonlinear Dynamics, School of Physics,  Bharathidasan University, Tiruchirapalli - 620 024, India}
\ead{velan@cnld.bdu.ac.in}
\begin{abstract}
In this article, we construct loop soliton solutions and mixed soliton - loop soliton solution for the Degasperis-Procesi equation. To explore these solutions we adopt the procedure given by Matsuno \cite{mats}. By appropriately modifying the $\tau$-function given in the above paper we derive these solutions. We present the explicit form of one and two loop soliton solutions and mixed soliton - loop soliton solutions and investigate the interaction between (i) two loop soliton solutions in different parametric regimes and (ii) a loop soliton with a conventional soliton in detail.

\end{abstract}

\maketitle

\section{Introduction}
In this paper, we investigate loop soliton solutions and their dynamics in the Degasperis-Procesi (DP) equation \cite{Degas},
\begin{eqnarray}
u_t+3\kappa^3 u_x-u_{xxt}+4 u u_x=3u_xu_{xx}+u u_{xxx},
\label{a1}
\end{eqnarray}
where subscripts denote partial derivatives and $\kappa$ is a positive parameter. The integrability of the DP equation was proved by constructing a Lax pair, deriving an infinite sequence of conservation laws and the existence of a bi-Hamiltonian structure \cite{Degasperis}. Eq.({\ref{a1}}) arises in a hydrodynamical context\cite{Johnson}. Interestingly when $\kappa=0$, Eq. (\ref{a1}) admits peakon solutions which are of the form $ u = c e^{-|x-ct|}$, where $c$ is the velocity of the peakon. Consequently the $N$ peakon solutions of the DP equation were constructed and the dynamics of these special solutions were also studied by  Lundmark and Szmigielski \cite{Lundmark}. The prolongation algebra and the Hamiltonian operator of this equation was reported in \cite{wang}. Vakhnenko and Parkes have studied various travelling wave solutions of the DP equation including hump-like, loop-like and coshoidal periodic-wave solutions \cite{Vakhnenko}. Qiao has come-up with  three new types of soliton solutions to this model, namely M-Shape peakons, dehisced solitons, double dehisced 1-peak solitons \cite{Qiao}.

 Here we are interested in exploring loop soliton solutions exhibited by Eq. (\ref{a1}). The motivation comes from the contemporary interest in studying loop soliton solutions in integrable nonlinear evolutionary equations \cite{VAVakhnenko, VOVakhnenko, AJMorrison, Morris, Sakovich, YMatsuno, BFFeng, Lin, Rogers, Matsuno, Matsunoy}. To our knowledge periodic inverted loop solutions and one loop soliton solution were reported for the DP equation \cite{Vakhnenko}. The explicit form of two loop soliton or higher order loop soliton solutions are not yet reported. Moreover the collision dynamics between the loop soliton solutions are also yet to be studied for this equation.

 To explore loop soliton solutions in Eq. (\ref{a1}) we follow the procedure given by Matsuno \cite{mats}. The author has derived $N$-soliton solutions for the DP equation from the modified version of the Kaup equation since the $\tau$-function of the latter is already known. By carefully examining the $\tau$- function we observe that one has the freedom in changing the sign of the coefficients of the exponential functions so that even after the sign changes the resultant $\tau$-function satisfies the bilinear identities. By taking this advantage, we reshape the $\tau$-function appropriately and derive loop soliton solutions in a systematic manner. In the case of a one soliton solution, since there is only one exponential function present in the $\tau$-function we simply change the sign of the coefficient in front of this exponential function and obtain a one loop soliton solution. In the case of two soliton solutions since there are two exponential functions present in the $\tau$-function one has two choices in fixing the coefficients of these exponential functions. Either the coefficients of both the exponential functions are negative or one coefficient is negative and the other one is positive.

 For the first choice we obtain two inverted loop solitons and for the second choice we obtain a mixed loop soliton - smooth soliton type solution. We consider both solutions and study their interaction properties. To begin with, we allow two loop solitary waves (among the two, one is longer and the other is shorter) to travel in a particular direction. As expected the taller loop wave travels faster than the smaller one and crosses the shorter loop wave in a finite time. When it approaches the smaller loop they start to interact. As a consequence the amplitude of the taller loop becomes shorter while the amplitude of the shorter loop becomes larger. This continues until the amplitudes of both the taller and shorter loop solitons become equal.   

In the second parametric regime, when the loop solitary waves cross each other the smaller and larger loop solitary waves overlap each other besides changing their amplitudes. In the third parametric regime we show that while the larger loop soliton overtakes the smaller one, the smaller loop solitary wave revolves circularly (in the clockwise direction) inside the larger loop solitary wave. 

 In the second case (mixed loop soliton - smooth soliton) we allow the loop solitary wave to interact with the smooth solitary wave. Here we bring out a totally different kind of interaction. The smaller loop solitary wave travels along the surface of the smooth soliton. All these results are new. 

 The plan of the paper is as follows. In Sec. 2, we recall the method of finding soliton solutions to this model. We present the method of constructing loop soliton solutions to this equation and derive the explicit expressions of one and two loop soliton solutions in Sec. 3. We investigate the collision dynamics, in the case of two loop solitons in detail. In Sec. 4, we derive mixed soliton - loop soliton solutions and study their interaction properties. We present our conclusions in Sec. 5.

\section{Method of finding soliton solutions \cite{mats}}
\label{sec:1}
In this section, we briefly recall the method of constructing $N$- soliton solutions of (\ref{a1}). The DP equation can be written in a compact form, $q_{\tilde {t}}+{q}^2 u_y=0$, first by defining a new variable $q$ with
\begin{eqnarray}
q^3 = u-{u_x}_x + \kappa^{3}
\label{b1}
\end{eqnarray}
and then introducing a reciprocal transformation, $dy = q dx - q u dt$, $d\tilde{t}=dt$, in the resultant equation. Rewriting the relation (\ref{b1}) one can express the old variable ($u$) in terms of new variable ($q$) as
\begin{eqnarray}
u=-q(\ln q)_{ty}+{q}^3-{\kappa}^3,
\label{b2}
\end{eqnarray}
where we have dropped the tilde above `$t$' for simplicity. From (\ref{b2}) we can find $u_{y}$. 
 Substituting this derivative in the expression $q_{t}+{q}^2 u_y=0$ and differentiating the resultant equation with respect to $y$ we obtain a third order ordinary differential equation in $q$. By successive differentiations this third order equation can be transformed to the first negative flow in the KK hierarchy (for more details one may refer \cite{mats}). The soliton solutions of this member can be identified from the soliton solutions of the modified version of the Kaup equation. From the known $\tau$-function of the Kaup equation one can go back and construct $N$-soliton solutions for the DP equation. To do this first one should express $q$ in terms of $\tau$-function of the Kaup equation. Doing so we get 
\begin{eqnarray}
{q}^2=-(\ln f)_{ty}+\kappa^2, \quad f=\mathrm{det}A,
\label{b3}
\end{eqnarray}
where 
 $A=(a_{jk})$ is a $2N\times2N$ matrix with elements
\numparts
\begin{eqnarray}
a_{jk}=(1+e^{\tilde\xi_j})\delta_{jk}+\frac{\tilde p_j-\tilde q_j}{\tilde p_j-\tilde q_k}(1-\delta_{jk}),\quad j,k=1,2,...,N
\label{b4}
~\\ \tilde\xi_{2j-1}=\tilde\xi_{2j}=k_j(y+\tilde c_j t-y_{j0})+\ln a_j, \quad j=1,2,...,N
\label{b4a} 
~\\ \tilde p_{2j-1}=q_j, \quad \tilde q_{2j-1}=-p_j, \quad \tilde p_{2j}=p_j, \quad \tilde q_{2j}=-q_j, \quad j=1,2,...,N,
\label{b4b}
\end{eqnarray}
\endnumparts
with the soliton velocity defined by 
\begin{eqnarray}
c_j=\frac{3 \kappa^4}{\kappa^2 k_j^2-1},
\quad a_j=\sqrt\frac{1-\frac{\kappa^2 k_j^2}{4}}{1-\kappa^2 k_j^2}, \quad j=1,2,...,N.
\label{b6}
\end{eqnarray}
 In the above $k_{j}$'s are wave parameters and the amplitude parameters $p_j$ and $q_j$ are found to be 
\begin{eqnarray}
p_j=\frac{k_j}{2}\bigg[1+\frac{2}{\kappa k_j}{\sqrt{\frac{1}{3}(1-\frac{\kappa^2 k_j^2}{4})}}}\bigg], 
\quad  q_j=\frac{k_j}{2}\bigg[1-\frac{2}{\kappa k_j}{\sqrt{\frac{1}{3}(1-\frac{\kappa^2 k_j^2}{4})}\bigg].
\label{b7}
\end{eqnarray}
In a nutshell, to construct the solutions, one should first get $q$ by substituting the explicit form of $f$ (vide Eq. (\ref{b4})) in (\ref{b3}). By plugging this $q$ in (\ref{b2}) one can get an implicit form of $u(y,t)$. Finally, substituting the expression $q$ in the mapping function,
\begin{eqnarray} 
x=\frac{y}{\kappa}+\int_{-\infty}^{y}(\frac{1}{q}-\frac{1}{\kappa}) dy + d,
\label{b8}
\end{eqnarray}
where $d$ is an integration constant, and integrating it one can get a relationship between $x$ and $y$.  We note that the function $q(y,t)$ should not vanish for any $y$ ant $t$. Otherwise a simple zero of $q$ would yield a logarithmic singularity of $x(y,t)$. Expressions (\ref{b2}) and (\ref{b8}) constitute the solution for the DP equation in parametric form. Since $q$ involves the variable $y$, after the integration, it is often difficult to express $y$ in terms of $x$ explicitly. In Ref.\cite{mats}, using the above procedure, the author has explicitly derived one and two soliton solutions of the DP Eq. (\ref{a1}).  

\section{\bf Method of finding loop solitons}
\label{sec:2}
To explore $N$-loop soliton solutions of the DP equation we modify the bilinear solution (\ref{b4}) as
\begin{eqnarray}
a_{jk}=(1-e^{\tilde\xi_j})\delta_{jk}+\frac{\tilde p_j-\tilde q_j}{\tilde p_j-\tilde q_k}(1-\delta_{jk}),\quad j,k=1,2,...,N
\label{c1}
\end{eqnarray}
with all other parameters $\tilde p_{j}$, $\tilde q_{j}$ and $\tilde q_{k}$  as given in Eqs. (\ref{b4a})-(\ref{b4b}). One can unambiguously prove the resultant $\tau$-function which comes out from the modified matrix elements also satisfies the bilinear identities. A simple sign change in the $\tau$-function leads to a new class of soliton solutions as we see below.
\subsection{One loop soliton solution}
\label{ssec:1}
Let us take $N=1$ in (\ref{b3}) with the matrix elements defined in (\ref{c1}). Expanding the determinant (\ref{b3}) we find the $\tau$-function modified to 
\begin{eqnarray}
f={a_1^2}(1-\frac{2}{a_1}e^{{\xi}_1}+e^{2{\xi}_1}),
\label{d1}
\end{eqnarray}
where
\begin{eqnarray}
\quad {\xi}_1=k_1(y+\frac{3\kappa^4}{\kappa^2 k_1^2-1}t-y_{10}),\quad a_1=\sqrt{\frac{1-\frac{\kappa^2 k_1^2}{4}}{1-\kappa^2 k_1^2}}, \quad (\kappa k_1>2).
\label{d2}
\end{eqnarray}
 One may observe that the sign in front of the second exponential in $f$ is positive in the case of one smooth soliton solution \cite{mats}. However, this change of sign has other impacts on the solution. For example, in the regime $\kappa k_{1}\le2$, $q$ has a simple zero and $u$ has a simple pole and the variable change $y$ has a logarithmic singularity, as we see below. 

 Substituting (\ref{d1}) into (\ref{b3}) and carrying out the derivatives and simplifying the resultant equation, we find 
\begin{eqnarray}
q=\kappa\frac{\cosh\xi_1-2 a_1+\frac{1}{a_1}}{\cosh\xi_1-\frac{1}{a_1}}.
\label{d3}
\end{eqnarray}
Now plugging $q$ and its derivatives in (\ref{b2}) and simplifying the resultant expressions to a compact form we arrive at
\begin{eqnarray}
u=-\frac{8\kappa^3}{a_1}\frac{(a_1^2-1)(a_1^2-\frac{1}{4})}{\cosh\xi_1-2 a_1+\frac{1}{a_1}}.
\label{d4}
\end{eqnarray}
\begin{figure}[!ht]     
\begin{center}
\includegraphics[width=0.85\linewidth]{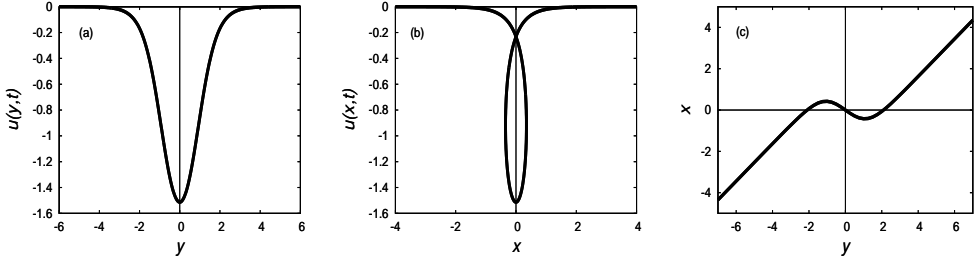}
\end{center}
\caption{Plots of (a) the inverted smooth solitary wave profile drawn $u(y,t)$ versus $y$ for the values $\kappa=1.1,k_1=2.1$, (b) the one loop solitary wave drawn $u(x,t)$ versus $x$ with the same parametric values and (c) the inverse coordinate transformation between $x$ versus $y$.}
\label{fig1}
\end{figure}

 Substituting $q(y,t)$ in Eq.(\ref{b8}) and performing the integration we obtain the coordinate transformation between $x$ and $y$ in the following implicit form
\begin{eqnarray}
x=\frac{y}{\kappa}-\ln\left[\frac{1+\alpha_1+(1-\alpha_1)e^{\xi_1}}{1-\alpha_1+(1+\alpha_1)e^{\xi_1}}\right],
\label{d6}
\end{eqnarray} 
where
\begin{eqnarray}
\alpha_1=\sqrt\frac{(2 a_1-1)(a_1+1)}{(2 a_1+1)(a_1-1)}.
\label{d7}
\end{eqnarray}
Since the variable $y$ depends on $\xi_{1}$ one cannot express $y$ in terms of $x$ explicitly. 

 Due to the coordinate transformation and the consequence of the change of sign in front of the exponential in the expression (\ref{d1}), the smooth solitary wave form morphed into a multivalued inverted loop like wave form with the condition given in Eq.(\ref{d2}). We present the nature of the obtained solutions in Fig. 1. To begin with, we draw the solution (\ref{d4}) in the variable $y$ where we get a smooth inverted solitary wave (Fig. 1a). While we depict the solution in terms of the original variable $x$ we obtain only the loop solitary wave form (Fig. 1b). To understand the geometrical connection between $x$ and $y$ we also draw a graph between these variables separately (Fig. 1c).
\subsection{Two loop soliton solutions}
\label{ssec:2}

 The $\tau$-function for the two loop soliton solution follows from Eq. (\ref{b3}) by restricting to $N=2$. The associated determinant of $f$ reads now,
\begin{equation}
 f = \left|
\begin{array}{cccc}
    1 - a_1 e^{\xi_1} & \frac{p_1+q_1}{2q_1} & \frac{p_1+q_1}{p_2+q_1} & \frac{p_1+q_1}{q_1+q_2}\\ 
    \frac{p_1+q_1}{2p_1} & 1-a_1e^{\xi_1} & \frac{p_1+q_1}{p_1+p_2} & \frac{p_1+q_1}{p_1+q_2}\\ 
    \frac{p_2+q_2}{p_1+q_2} & \frac{p_2+q_2}{q_1+q_2} & 1-a_2e^{\xi_2} & \frac{p_2+q_2}{2q_2}\\ 
    \frac{p_2+q_2}{p_1+p_2} & \frac{p_2+q_2}{p_2+q_1} & \frac{p_2+q_2}{2p_2} & 1-a_2e^{\xi_2}
\end{array} \right|.
\label{e0}
\end{equation}
When compared to two smooth soliton solutions the signs in front of the parameters $a_1$ and $a_2$ are different \cite{mats}. Substituting the expressions of  $p_j$'s and $q_j$'s, $j=1,2$, given in (\ref{b7}), in the above determinant and expanding it we obtain the explicit form of the $\tau$-function as
\begin{eqnarray}
f=(a_1 a_2)^2\bigg(\delta^2-\frac{2\delta}{a_1}e^{\xi_1}-\frac{2\delta}{a_2}e^{\xi_2}+e^{2\xi_1}+e^{2\xi_2}+\frac{2\nu}{a_1 a_2}e^{\xi_1+\xi_2}
    -\frac{2}{a_2}e^{2\xi_1+\xi_2} \nonumber \\ \qquad -\frac{2}{a_1}e^{\xi_1+2\xi_2}+e^{2(\xi_1+\xi_2)}\bigg),
\label{e1}
\end{eqnarray}
where
\begin{eqnarray}
\nonumber \delta=\frac{(k_1-k_2)^2[\kappa^2(k_1^2-k_1 k_2 +k_2^2)-3]}{(k_1+k_2)^2[\kappa^2(k_1^2+k_1 k_2 +k_2^2)-3]},
 \label{e2}
\end{eqnarray}
\begin{eqnarray}
\nu=\frac{(2k_1^4-k_1^2 k_2^2+2 k_2^4)\kappa^2-6(k_1^2+k_2^2)}{(k_1+k_2)^2[\kappa^2(k_1^2+k_1 k_2 +k_2^2)-3]}.
\label{e3}
\end{eqnarray}

 Substituting (\ref{e1}) and (\ref{e2}) in equation (\ref{b3}) we obtain a compact expression for $q(y,t)$, that is
\begin{eqnarray}
q(y,t)=\kappa\frac{g}{\tilde f}=\kappa \frac{g_{1}g_{2}}{f/(a_1 a_2)^2}.
\label{e4}
\end{eqnarray}
The explicit forms of $g_1$ and $g_2$ are given by 
\begin{eqnarray}
\hspace{-1.9cm}g_1=\delta-\frac{2-\kappa k_1}{2 a_1(1+\kappa k_1)}e^{\xi_1}-\frac{2-\kappa k_2}{2 a_2(1+\kappa k_2)}e^{\xi_2}+\frac{(2-\kappa k_1)(2-\kappa k_2)}{4 a_1 a_2(1+\kappa k_1)(1+\kappa k_2)}e^{\xi_1+\xi_2}
\label{e5}
\end{eqnarray}
\begin{eqnarray}
\hspace{-1.9cm}g_2=\delta-\frac{2+\kappa k_1}{2 a_1(1-\kappa k_1)}e^{\xi_1}-\frac{2+\kappa k_ 2}{2 a_2(1-\kappa k_2)}e^{\xi_2}+\frac{(2+\kappa k_1)(2+\kappa k_2)}{4 a_1 a_2(1-\kappa k_1)(1-\kappa k_2)}e^{\xi_1+\xi_2}.
\label{e6}
\end{eqnarray}
We note here that the right hand side of Eq.(\ref{b3}) turns out to be a perfect square and the numerator in the resultant expression is factorized into the product of two functions, that is $g=g_{1}g_{2}$ with $g_1$ and $g_2$ being polynomials of $e^{\xi_1}$ and $e^{\xi_2}$. 

 Substituting (\ref{e4}) in (\ref{b2}) and performing the differentiation we find the solution $u(y,t)$ be of the form 
\begin{eqnarray}
u=\kappa ^3 \frac{h}{g},
\label{e7} 
\end{eqnarray}
with
\begin{eqnarray}
h=-{\frac{9\delta\kappa^2 k_1^2}{a_1(1-\kappa^2 k_1^2)^2}e^{\xi_1}+\frac{9\delta\kappa^2 k_2^2}{a_2(1-\kappa^2 k_2^2 )^2}e^{\xi_2}}+\frac{9\kappa^2 k_1^2}{a_1(1-\kappa^2 k_1^2)^2}e^{\xi_1+2\xi_2} \\ \nonumber \qquad+\frac{9\kappa^2 k_2^2}{a_2(1-\kappa^2 k_2^2)^2}e^{2\xi_1+\xi_2}.
\label{e8}
\end{eqnarray}
We observe that the numerator of $u$ is of the same form as for the two soliton solutions with the only difference in an over all sign change. The coordinate transformation between $x$ and $y$ is given by
\begin{eqnarray}
x(y,t)=\frac{y}{\kappa}+\ln(\frac{g_1}{g_2})+d,
\label{e9}
\end{eqnarray}
where $g_1$ and $g_2$ are given in (\ref{e5}) and (\ref{e6}).
\begin{figure}[!ht]     
\begin{center}
\includegraphics[width=0.85\linewidth]{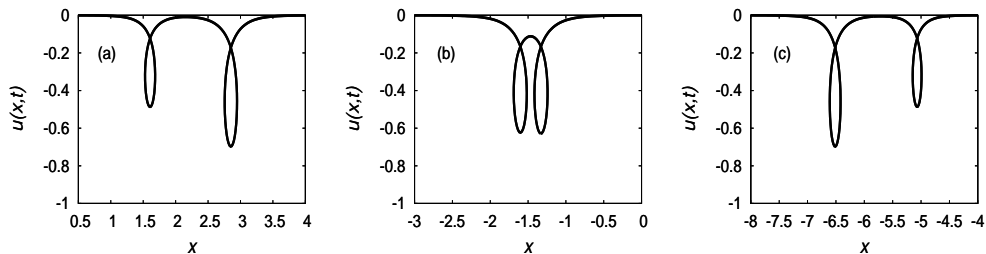}
\end{center}
\caption{ Two loop soliton solutions of (\ref{a1}) drawn at different times, (a) $t=-10$, (b) $t=-0.8$ and (c) $t=10$ with $\kappa=1.5,k_1=3.2,k_2=3.8$ and $d=0$.}
\label{fig2}
\end{figure}
\subsection{Two loop soliton interactions}
\label{ssec:3}
Expressions (\ref{e7}) and (\ref{e9}) provide the complete description of the two loop soliton solution in the form of the parametric representation.

It describes the two loop solitary wave troughs. Fig. (2) shows the interaction between two loop solitary wave troughs. Here we plot the solution $u$ in terms of $x$ with the parameter values $\kappa=1.5,k_1=3.2,k_2=3.8$ and $d=0$. Both the solitary waves propagate towards the negative $x$ direction. Initially  at $t=-10$, the larger loop solitary wave is well separated from the smaller one (Fig. 2a). Since the larger loop soliton travels faster than the smaller one it starts  to cross the smaller one in a finite time. Now the larger wave loses energy to the smaller wave. As a consequence the amplitude of the larger loop decreases and the amplitude of the smaller one increases.  At $t=-0.8$, these two loop solitons have equal amplitude or energy (Fig. 2b). The two loop solitons do not superpose into a single wave as one can see from (Fig. 2b). After the interaction (elastic interaction) these two solitons re-emerge and travel in their original direction by keeping their original amplitudes (Fig. 2c).  
  
\begin{figure}[!ht]     
\begin{center}
\includegraphics[width=0.85\linewidth]{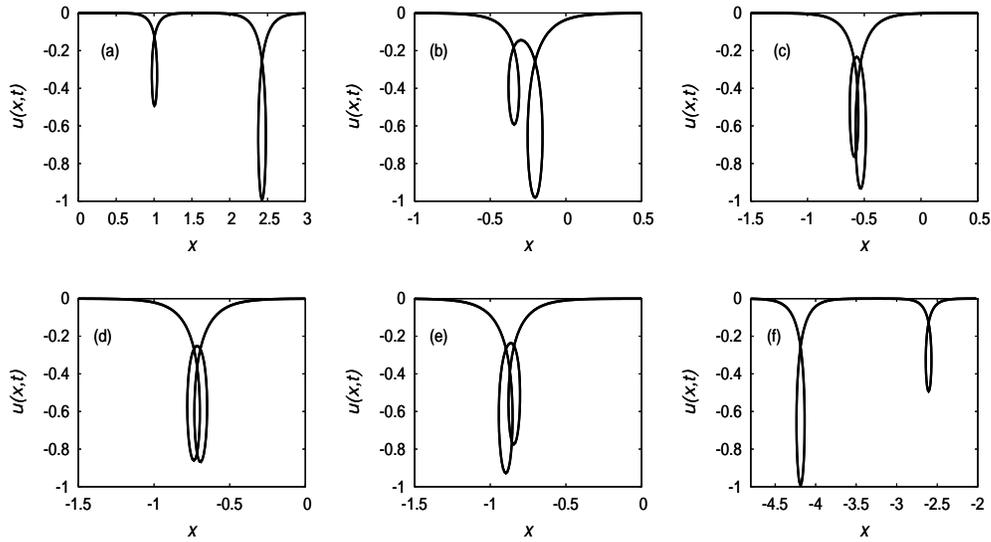}
\end{center}
\caption{ The demonstration of two anti-loop soliton overlapping interactions in (\ref{a1}).}
\label{fig3}
\end{figure}

  Now we investigate the two loop soliton interaction in a different parametric regime, say for example $\kappa=2.5,k_1=3.4,k_2=4.8$ and $d=0$. To begin with ($t=-5$) the two loop solitons are well separated (Fig. 3a). As in the previous case both the larger and smaller loop waves move towards left. At $t=-1$ the larger loop solitary  wave approaches the smaller one (Fig. 3b).  Fig. 3c shows the changes in their amplitudes when they start to interact with the amplitude of the taller one becoming shorter and the smaller one becoming larger. Interestingly at $t=-0.25$ the amplitudes of both the taller and smaller waves becomes equal and overlap each other (Fig. 3d). As time progresses the two loop solitons emerge out from each other in the original direction by changing their heights (Fig. 3e). At $t=5$ the loops are completely separated and attain their original amplitudes.

 Another interesting loop soliton interaction can be seen by fixing the parametric values  as $\kappa=3.5,k_1=10.4,k_2=4.2$ and $d=0$. We draw the interaction picture in Fig. 4. The two loop solitons are well separated from each other in the beginning, see Fig. 4a. When time moves on the larger loop meets the smaller loop (Fig. 4b) and they start to interact (Fig. 4c). In distinction to the previous two cases the smaller loop revolves inside the larger loop in the clockwise direction (Fig. 4c-4e) and finally emerges out from the larger loop. Later both the loops attain their original amplitudes.
\begin{figure}[!ht]     
\begin{center}
\includegraphics[width=0.85\linewidth]{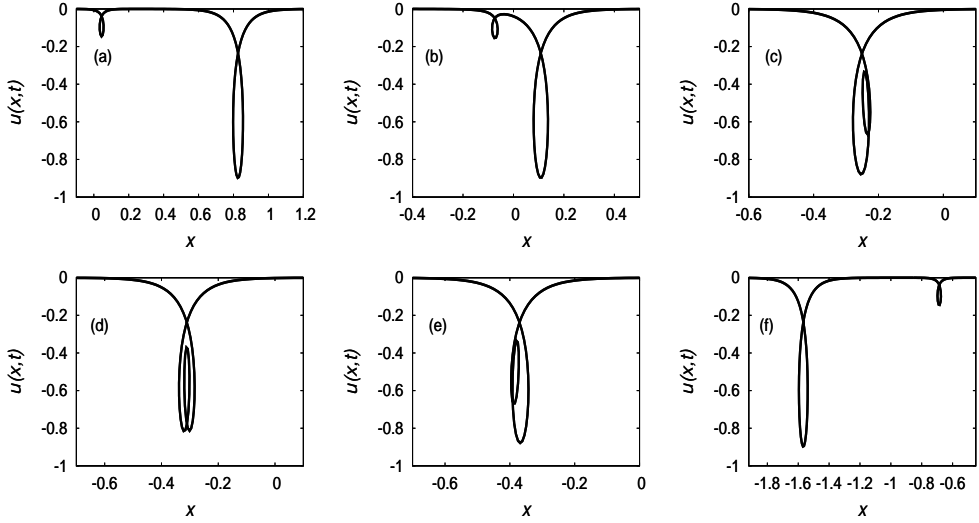}
\end{center}
\caption{ The demonstration of clockwise rotation of a smaller loop soliton inside a bigger loop soliton.}
\label{fig4}
\end{figure}

\section{\bf Mixed  loop soliton - soliton solution of the DP equation}
\label{sec:3}
In the previous section, while deriving the two loop soliton solutions we changed the sign of the coefficients of both exponential functions in the $\tau$-function to negative values, see Eq. (\ref{e0}). In this section, we consider another possibility, we change the sign of one exponential function only 
\begin{equation}
f = \left|
\begin{array}{cccc}
    1 - a_1 e^{\xi_1} & \frac{p_1+q_1}{2q_1} & \frac{p_1+q_1}{p_2+q_1} & \frac{p_1+q_1}{q_1+q_2}\\ 
    \frac{p_1+q_1}{2p_1} & 1-a_1e^{\xi_1} & \frac{p_1+q_1}{p_1+p_2} & \frac{p_1+q_1}{p_1+q_2}\\ 
    \frac{p_2+q_2}{p_1+q_2} & \frac{p_2+q_2}{q_1+q_2} & 1 + a_2e^{\xi_2} & \frac{p_2+q_2}{2q_2}\\ 
    \frac{p_2+q_2}{p_1+p_2} & \frac{p_2+q_2}{p_2+q_1} & \frac{p_2+q_2}{2p_2} & 1 + a_2e^{\xi_2}
\end{array} \right|. 
\label{f0}
\end{equation}
 For the opposite sign choice we would get inverted solutions. By performing the same procedure as in sub-section 3.1, we obtain the following expression for the functions $f$, $q$ and $u$ respectively, that is 
\begin{eqnarray}
f=(a_1 a_2)^2\bigg(\delta^2-\frac{2\delta}{a_1}e^{\xi_1}+\frac{2\delta}{a_2}e^{\xi_2}+e^{2\xi_1}+e^{2\xi_2}-\frac{2\nu}{a_1 a_2}e^{\xi_1+\xi_2}
    +\frac{2}{a_2}e^{2\xi_1+\xi_2}\nonumber \\ \qquad -\frac{2}{a_1}e^{\xi_1+2\xi_2}+e^{2(\xi_1+\xi_2)}\bigg),
\label{f1}
\end{eqnarray}
\begin{eqnarray}
q(y,t)=\kappa\frac{g}{\tilde f}=\kappa \frac{g_{1}g_{2}}{f/(a_1 a_2)^2},
\label{f2}
\end{eqnarray}
\begin{eqnarray} 
u=\kappa ^3 \frac{h}{g},
\label{f5}
\end{eqnarray}
where,
\begin{eqnarray}
\hspace{-1.9cm}g_1=\delta-\frac{2-\kappa k_1}{2 a_1(1+\kappa k_1)}e^{\xi_1}+\frac{2-\kappa k_2}{2 a_2(1+\kappa k_2)}e^{\xi_2}-\frac{(2-\kappa k_1)(2-\kappa k_2)}{4 a_1 a_2(1+\kappa k_1)(1+\kappa k_2)}e^{\xi_1+\xi_2}
\label{f3}
\end{eqnarray}
\begin{eqnarray}
\hspace{-1.9cm}g_2=\delta-\frac{2+\kappa k_1}{2 a_1(1-\kappa k_1)}e^{\xi_1}+\frac{2+\kappa k_2}{2 a_2(1-\kappa k_2)}e^{\xi_2}-\frac{(2+\kappa k_1)(2+\kappa k_2)}{4 a_1 a_2(1-\kappa k_1)(1-\kappa k_2)}e^{\xi_1+\xi_2}\\ \nonumber
\label{f4}
\end{eqnarray}
\begin{eqnarray} 
\hspace{-1.9cm} h={-\frac{9\delta\kappa^2 k_1^2}{a_1(1-\kappa^2 k_1^2)^2}e^{\xi_1}+\frac{9\delta\kappa^2 k_2^2}{a_2(1-\kappa^2 k_2^2 )^2}e^{\xi_2}}-\frac{9\kappa^2 k_1^2}{a_1(1-\kappa^2 k_1^2)^2}e^{\xi_1+2\xi_2} \\ \nonumber  \qquad+\frac{9\kappa^2 k_2^2}{a_2(1-\kappa^2 k_2^2)^2}e^{2\xi_1+\xi_2}.
\label{f6}
\end{eqnarray}
Eq. (\ref{f5}) is a mixed soliton-loop  soliton solution of the DP equation with the coordinate transformation given in Eq. (\ref{e8}).
\subsection{Mixed soliton - loop soliton interaction}
\label{ssec:4}
In this sub-section we investigate the head on collision between a loop soliton with a smooth soliton. Eq. (\ref{f5}) describes a smooth solitary wave form travelling towards right and a loop wave trough travelling towards left. These two waves propagate in opposite directions and collide with each other. To investigate the outcome we draw the solution (\ref{f5}) in Fig. (5) with the parameter values $\kappa=0.91,k_1=2.6$ and $k_2=0.91$. In this case we observe that the anti-loop soliton creates a secondary wave crest. As time goes on the secondary pulse grows in amplitude while the primary wave loses its momentum. One can observe a double peaked smooth wave when the loop wave reaches the top of the smooth solitary wave. After a certain time the smaller loop starts sliding in the left direction. This interaction process continues until the two waves separate from each other.

\begin{figure}[!ht]     
\begin{center}
\includegraphics[width=0.85\linewidth]{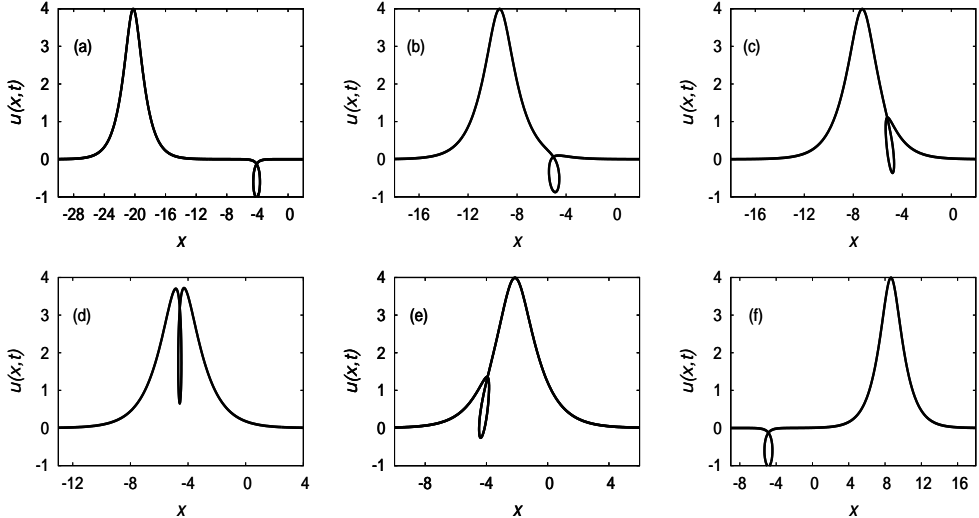}
\end{center}
\caption{ The demonstration of mixed loop soliton versus smooth soliton interaction in (\ref{a1}).}
\label{fig5}
\end{figure}

\section{\bf Conclusion}
\label{sec:4}
In this article, we have focused our attention on obtaining loop soliton solutions of the DP Eq. (\ref{a1}). We have recovered these solutions directly from the $\tau$- function of the modified version of the Kaup equation. We have given parametric representations for both the pure loop solitons and mixed loop - smooth solitons and studied their wave dynamics. We have investigated the loop - loop soliton and mixed loop - soliton interactions in detail. In the case of the loop soliton interaction, when the amplitudes of the loop solitons are dissimilar, we observed that the smaller loop travels across the larger one in three different fashions before being shifted. We have shown the formation of double peaked soliton waves in the case of mixed loop - smooth soliton interaction. Currently, we are formulating $N$- loop soliton solutions for the DP equation.

\section*{Acknowledgment}

The work forms part of a research project sponsored by the University Grants Commission (UGC), Government of India.

\end{document}